\documentclass[prb,twocolumn,showpacs,preprintnumbers,amsmath,amssymb,citeautoscript]{revtex4}

\usepackage{times}
\usepackage{graphicx}
\usepackage{dcolumn}
\usepackage{bm}

\begin{document}

\title{Nodal gap structure of BaFe$_2$(As$_{1-x}$P$_x$)$_2$ from angle-resolved thermal conductivity in a magnetic field}

\author{M.~Yamashita$^1$}
\author{Y.~Senshu$^1$}
\author{T.~Shibauchi$^1$}
\author{S.~Kasahara$^2$}
\author{K.~Hashimoto$^1$}
\author{D.~Watanabe$^1$}
\author{H.~Ikeda$^1$}
\author{T.~Terashima$^2$}
\author{I.~Vekhter$^3$}
\author{A.\,B.~Vorontsov$^4$}
\author{Y.~Matsuda$^1$}

\affiliation{$^1$Department of Physics, Kyoto University, Kyoto 606-8502, Japan}
\affiliation{$^2$Research Center for Low Temperature and Materials Sciences, Kyoto University, Kyoto 606-8502, Japan}
\affiliation{$^3$Department of Physics and Astronomy, Louisiana State University, Baton Rouge, Louisiana, 70803, USA}
\affiliation{$^4$Department of Physics, Montana State University, Bozeman, Montana, 59717, USA}

\date{\today}

\begin{abstract}
The structure of the superconducting order parameter in the iron-pnictide superconductor BaFe$_2$(As$_{0.67}$P$_{0.33}$)$_2$  ($T_c=31$\,K) with line nodes is studied by the angle-resolved thermal conductivity measurements in a magnetic field rotated within the basal plane.  We find that the thermal conductivity displays distinct fourfold oscillations with minima when the field is directed at $\pm45^\circ$ with respect to the tetragonal $a$ axis.  We discuss possible gap structures that can account for the data, and conclude that the observed results are most consistent with the closed nodal loops located at the flat parts of the electron Fermi surface with high Fermi velocity.
\end{abstract}



\maketitle

The fundamental mechanism that causes high-temperature superconductivity is a central issue in the physics of Fe-pnictides. Knowledge of the superconducting (SC) gap structure is a major step toward identifying the interactions that produce pairing.  However, despite tremendous experimental and theoretical efforts, the SC gap structure remains one of the most controversial topics in these materials. The Fe-pnictides have several electron and hole bands and the interband interaction is very important for superconductivity \cite{Pag10}.  Nesting between electron and hole bands promotes antiferromagnetism and the $s_{\pm}$-wave gap symmetry with a sign change between nested bands \cite{Maz08,Kur09}.  In contrast, orbital fluctuations promote an $s_{++}$ gap without a sign change \cite{Kon10}.

Recent experiments revealed that several Fe-pnictide superconductors have line nodes in the SC gap. They include LaFePO ($T_c\approx6$\,K) \cite{Fle09,Yam09}, heavily hole-doped KFe$_2$As$_2$ ($T_c\approx 3$\,K) \cite{Has10b} and BaFe$_2$(As$_{1-x}$P$_x$)$_2$ \cite{Has10a,Nak10a}. The presence of line nodes in the latter material is especially intriguing because the transition temperature of this system in the optimally doped region, $T_c=31$\,K at $x=0.33$ \cite{Kas10}, is comparable to the highest $T_c$ of hole-doped (Ba$_{1-x}$K$_x$)Fe$_2$As$_2$ ($T_c=37$\,K) whose Fermi surface (FS) is fully gapped in the SC state \cite{Has09,Ding}.  In BaFe$_2$(As$_{1-x}$P$_x$)$_2$, substitution of isoelectronic P for As suppresses the magnetic order and induces superconductivity without charge doping.  This system is very clean as demonstrated by the observed quantum oscillations \cite{Shi10} and by the low critical current density in the vortex state.\cite{Beek} 
In addition, the $x$-dependence of several physical quantities is consistent with a quantum critical point near the boundary of magnetism and superconductivity~\cite{Nak10b}.  Therefore this system is suitable for studying the intrinsic nature of the high-$T_c$ superconductivity and for determining the detailed SC gap structure.

Here we report angle-resolved measurements of the thermal conductivity $\kappa$ in a rotated magnetic field {\boldmath $H$}, which is a bulk probe of the position of the gap nodes in unconventional superconductors \cite{Vek99,Mat06,Vor06,Gra08,Chu10,Wen10,Vor10}. We observed distinct fourfold modulations under field rotation within the $ab$ plane.  Combined with the results of the superfluid density $\rho_s$ \cite{Has10a}, specific heat $C$ \cite{Kim10}, the band structure calculation and guided by comparison with the theoretical predictions for angular variations of $\kappa$ for different FS shapes and nodal structures, we conclude that the nodal lines form closed loops at the flatter part of the outer electron pocket near the intersection of the electron pockets with the $\Gamma$-$X$ line.

\begin{figure}[t]
\includegraphics[width=0.97\columnwidth]{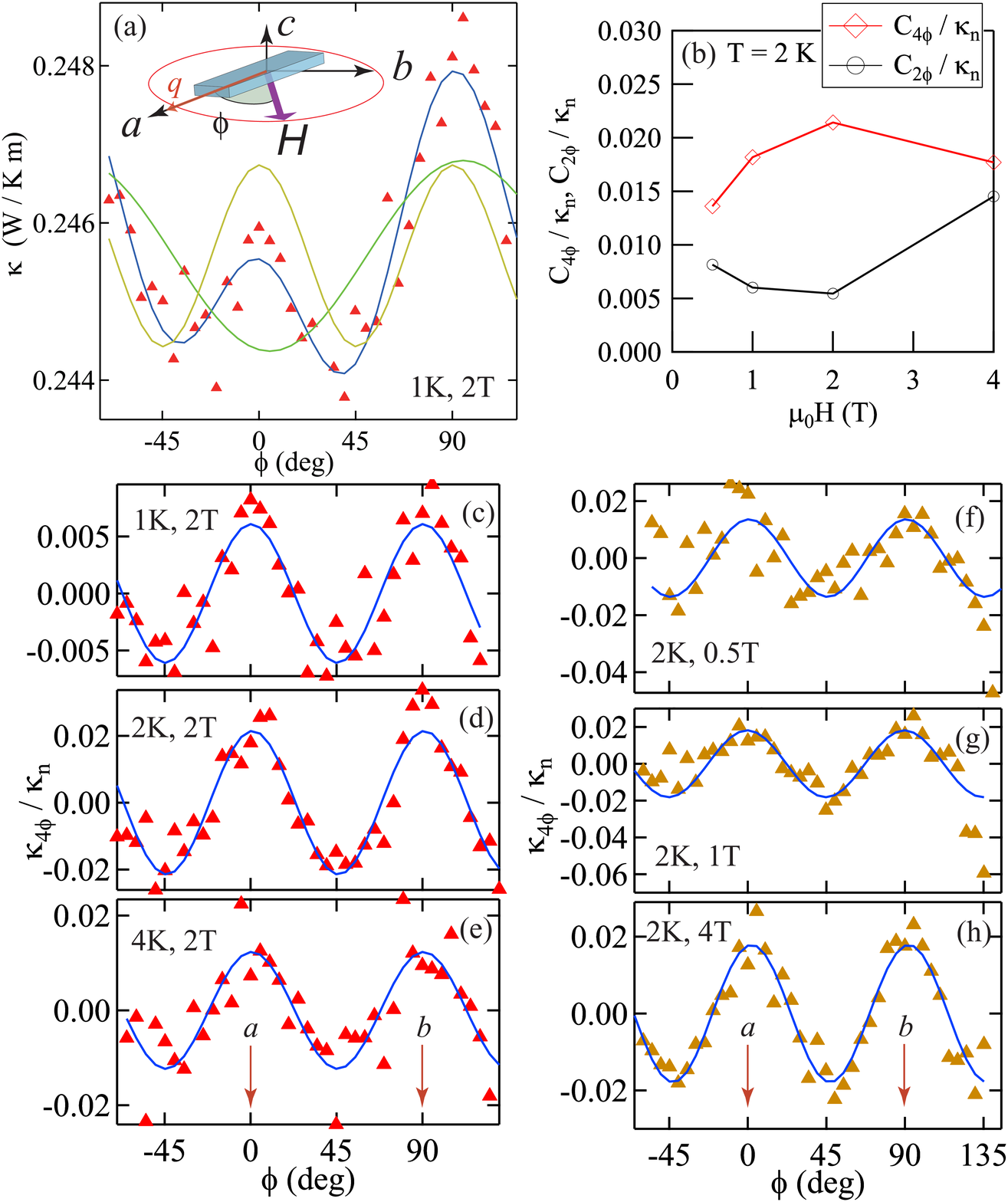}
\caption{
(Color online)
(a) Angular variation of $\kappa(\phi)$ in BaFe$_2$(As$_{0.67}$P$_{0.33}$)$_2$ with a fit to $\kappa(\phi)=\kappa_0+\kappa_{2\phi}+\kappa_{4\phi}$ (blue line).  The green (yellow) line represents the twofold (fourfold) term obtained by the fitting. The inset illustrates the measurement configuration.
(b) The magnitudes of $\kappa_{2\phi}$ and $\kappa_{4\phi}$ at 2\,K are plotted against the field magnitude $H$.
(c)--(e) The fourfold term $\kappa_{4\phi}$ at $\mu_0H=2$\,T. (f)--(h) $\kappa_{4\phi}$ at $T=2$\,K. }
\label{fig:kappa_phi}
\end{figure}

$\kappa$ was measured for BaFe$_2$(As$_{0.67}$P$_{0.33}$)$_2$ single crystals \cite{Kas10} by the steady-state method with the thermal current {\boldmath $q$} along the tetragonal $a$ axis [Fig.\:\ref{fig:kappa_phi}(a)]. We rotate {\boldmath $H$} within the $ab$ plane with a misalignment of less than 0.02$^{\circ}$ from the plane. The temperature and field dependence of $\kappa/T$ of the present crystal agree with the data in Ref.\:\onlinecite{Has10a}, including finite residual $\kappa/T$ at $T \rightarrow 0$\,K in zero field and $\sqrt{H}$ dependence of $\kappa/T$, expected of linear nodes.

Figure\:\ref{fig:kappa_phi}(a) depicts the angular variation of $\kappa(\phi)$, where $\phi$ is the angle between the $a$ axis and {\boldmath $H$}.  We measured $\kappa(\phi)$ by changing $\phi$ after field cooling at $\phi=0^{\circ}$.  The subsequent measurement with an inverted direction of rotation did not produce any hysteresis in $\kappa(\phi)$. At $\phi=\pm45^{\circ}$, $\kappa(\phi)$ shows minima, indicating the presence of fourfold modulation.  As shown by the solid lines, $\kappa(\phi)$ at each $T$ and $H$ can be decomposed as $\kappa(\phi)=\kappa_0+\kappa_{2\phi}+\kappa_{4\phi}$, where $\kappa_0$ is a constant, and $\kappa_{2\phi}=C_{2\phi}\cos2\phi$ and $\kappa_{4\phi}=C_{4\phi}\cos4\phi$ are the terms with twofold and fourfold symmetry respectively. The twofold contribution appears because of the difference in transport with the field parallel to and normal to the heat current \cite{Vor06}, and is present even for an isotropic gap. However, in contrast to most previously studied nodal superconductors \cite{Mat06}, the amplitude of the twofold term is smaller than that of the fourfold term [Fig.\:\ref{fig:kappa_phi}(b)]. As we see below, this imposes constraints on the shape of the superconducting gap.

Figures\:\ref{fig:kappa_phi}(c)--1(h) display $\kappa_{4\phi}/\kappa_n$ at low $T$ and low $H$, which is directly related to the SC gap structure.  The normal-state $\kappa_n$ value is obtained from the residual resistivity \cite{Kas10} by using the Wiedemann-Franz law. We observe clear fourfold oscillations with the amplitude $|C_{4\phi}|/\kappa_n$$\sim$0.01--0.02. Such oscillations are due to the creation of unpaired quasiparticles in the regions with small or vanishing gaps by the applied field. Previous $\kappa/T(T,H)$ and $\rho_s(T)$ measurements \cite{Has10a} already indicate the presence of nodes rather than the gap minima, so we focus solely on the location of the nodes. The strength of the pair breaking in a given near-nodal region vanishes if the Fermi velocity $\bm v_F$ is parallel to $\bm H$, and is maximal for $\bm v_F\perp\bm H$. At the lowest energies this can be explained by the Doppler shift of the quasiparticle energy by the amount $\delta\epsilon(\bm r, \bm v_F)=\bm v_F(\bm k)\cdot \bm p_s(\bm r)$, where $\bm p_s$ is the superflow due to the vortices in the plane normal to {\boldmath $H$} \cite{Vek99}, and more microscopic theories retain this structure \cite{Vor06}. Hence rotating $\bm H$ provides a probe of the node positions.

\begin{figure}[t]
\includegraphics[width=0.97\columnwidth]{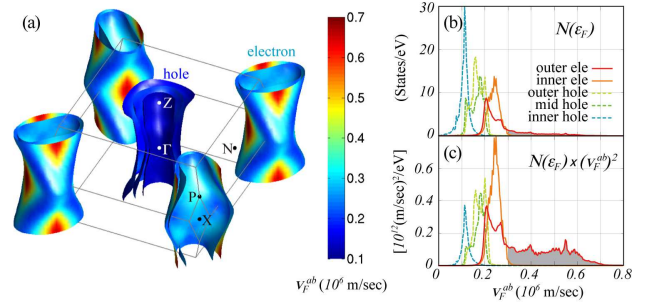}
\caption{(Color online) (a) Fermi surface of BaFe$_2$(As$_{0.67}$P$_{0.33}$)$_2$. The shading represents the in-plane Fermi velocity $v_F^{ab}$. 
Note that the flat parts of the outer electron sheets [near point A in Figs.\:3(a)--3(c)] have high $v_F^{ab}$ values, which correspond to the gray shading in Figs.\:2(b) and 2(c).
(b), (c) The DOS, $N(\varepsilon_F)~(\propto C/T)$ and $N(\varepsilon_F)(v_F^{ab})^2~(\propto \kappa/T)$ of each band as a function of $v_F^{ab}$. }
\label{fig:FS}
\end{figure}

\begin{figure}[t]
\begin{center}
\includegraphics[width=0.9\columnwidth]{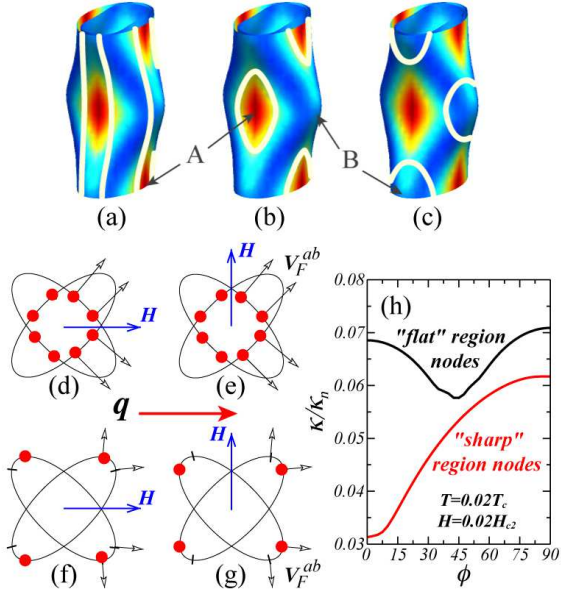}
\caption{
(Color online)
(a)--(c) Possible node positions on the electron Fermi surfaces. For $k_z=0$ and $\pm\frac{2\pi}{c}$, electron surface cross-sections are approximately elliptical, (d)--(g). (d), (e) When the nodes are on the flat part, the field-induced states near all nodes (filled red circles) have a comparable contribution to the heat current for $\bm H\parallel\bm q$ (d) and for $\bm H\perp \bm q$ (e), so that the twofold component of $\kappa$ is nearly absent. (f), (g) When the nodes are on the sharp ends, the field-induced DOS is small for nodes with ${\bm v_F}$ nearly parallel to ${\bm H}$ (black crosses). The nodes with more quasiparticles (red circles) have a different angle between ${\bm v_F}$ and $\bm q$ for the two field orientations, leading to a dominant two-fold anisotropy. (h) Comparison of $\kappa(\phi)$ for the two cases with the scattering rate $\gamma=\Gamma/2\pi T_c=0.01$, phase shift of scattering $\delta=60^\circ$, and the ratio of interband to intraband scattering $\delta v=0.9$.}
\label{fig:nodes2D}
\end{center}
\end{figure}

BaFe$_2$(As$_{0.67}$P$_{0.33}$)$_2$ is a multiband system. We obtained the FS from the band structure calculations including spin-orbit coupling using a density functional theory implemented in the \textsc{wien2k} code. The FS consists of five quasi-cylindrical pockets, three hole pockets at the center of the Brillouin zone, and two electron pockets centered at its corners [Fig.\:\ref{fig:FS}(a)]. Because of the body-centered tetragonal symmetry, the $k_z$ cuts of electron sheets are elliptical, with only twofold symmetry. These ellipses are rotated by 90$^\circ$ between $k_z$=0 and $\pm\frac{2\pi}{c}$, forming a so-called ``snake that swallowed a chain'' shape. As shown in Figs.\:\ref{fig:nodes2D}(a)--3(c), flat regions appear in the electron pockets around point A (along $\Gamma$-$X$) and the FS has a higher curvature near point B.

The first question is at which FS(s) are the nodes located. We believe that the nodes are on the electron sheets. First, recent bulk-sensitive laser angle-resolved photoemission spectroscopy (ARPES) experiments report isotropic gaps in all three hole pockets around $Z$ point \cite{Shimo11}. This by itself excludes the $d$-wave gap symmetry and we focus on the $A_{1g}$ representation. Second, there are significant changes in the shape of the hole FSs with $x$ in BaFe$_2$(As$_{1-x}$P$_x$)$_2$, while the shape of the electron surfaces remains nearly the same \cite{Kas10}. The slope of the $T$-linear penetration depth, indicative of robust line nodes, simply scales with $T_c$ in a wide range of $x$ \cite{Has11}, implying little change in $v_F$ at the near-nodal regions. Finally, in an applied field the residual $\kappa/T$ exhibits a clear $\sqrt{H}$ dependence \cite{Has10a} over a much wider range than the residual specific heat $C/T$, which looks nearly linear above 5\,T \cite{Kim10}. This suggests that the nodes are located in regions with higher in-plane Fermi velocity $v_F^{ab}$, as those give a lower relative contribution to $C/T$ [proportional to the density of states, DOS, $N(\varepsilon_F, \bm H)$] than to transport [proportional to $N(\varepsilon_F)(v_F^{ab})^2$]. The shading of Fig.\:\ref{fig:FS}(a) represents the magnitude of $v_F^{ab}$ with the red (blue) area denoting regions with high (low) Fermi velocity. It is clear that the hole pockets have low velocity and heavy mass quasiparticles, while the electron FS sheets have high velocity and light mass. Considering that the electron pockets have a lower scattering rate than the hole pockets, their contribution to the total thermal conductivity is even more pronounced. Note that these regions with high $v_F^{ab}$ also have a higher Doppler shift $\delta\varepsilon$ for $\bm H\perp ab$. Together, these observations strongly support the scenario with the nodes on the electron sheet.

The above analysis also suggests that the nodes may be located in the flat parts in the outer electron sheets, where the Fermi velocity is highest. Indeed, Figs.\:\ref{fig:FS}(b), and 2(c) depict  $N(\varepsilon_F)$ ($\propto C/T$) and $N(\varepsilon_F)(v_F^{ab})^2$ ($\propto\kappa/T$) as a function of $v_F^{ab}$.  The gray shaded areas indicate the high velocity regions with $v_F^{ab}\geq 0.3 \times 10^6$\,m/s, flat parts in the outer electron sheets.  Although their contribution is less than 5\% in the total DOS, it exceeds 30\% in the net $\kappa/T$, or more if the longer mean free path is accounted for.


The profile of $\kappa(\phi)$ under a rotated field allows us to further explore the nodal locations. The fourfold oscillations of $\kappa(\phi)$ of similar amplitude to that in materials with vertical line nodes argue against nearly horizontal line nodes suggested by several theories \cite{Gra10,Suz11}. Depending on the temperature and field range either minima or maxima of $\kappa_{4\phi}$ can indicate nodal directions \cite{Vor06}. The inversion of the oscillations reported in Fe(Se,Te) with a similar FS structure \cite{Wen10} occurs at a much higher temperature and field than our range (0.03$\leq T/T_c \leq$0.13 and 0.007$\alt H/H_{c2} \alt $0.06). 
Therefore the minima at $\phi=\pm45^{\circ}$ observed in our low-$T$ and low-$H$ range indicate that the nodes are located at the position of the FS where ${\bm v_F}$ is nearly parallel to [$\pm$1, $\pm$1, 0] directions.

In the $A_{1g}$ symmetry we consider three possible nodal structures shown in Figs.\:\ref{fig:nodes2D}(a)--3(c); (a) eight vertical line nodes, (b) closed loop line nodes in the flat part \cite{Maz10}, and (c) nodal loops in the high curvature part.
To distinguish between those possibilities we compute the profile of the thermal conductivity as a function of the field angle using the microscopic approach of Ref.\:\onlinecite{Vor06} generalized to the multiband system in Ref.\:\onlinecite{Vor10}, which has been successful in describing $C/T$ and $\kappa/T$ oscillations on equal footing. We take one electron and one hole FS, assume the hole FS to be fully gapped, and consider a nodal order parameter on the electron sheet. An intuitive physical analysis connects the smallness of the twofold term with the nodes in the flat part of the ellipse---see Figs.\:\ref{fig:nodes2D}(d)--3(g). The two ellipsoids represent the electron sheets at $(\frac{\pi}{a},\frac{\pi}{a})$ and $(\frac{\pi}{a},-\frac{\pi}{a})$ points respectively. 
Recall that quasiparticles are predominantly generated at locations where $\bm v_F\perp \bm H$ (filled red circles), and quasiparticles with $\bm v_F\parallel \bm q$ contribute more to the heat transport. If the nodes are in the high curvature region [in the case of Fig.\:\ref{fig:nodes2D}(c)], the Fermi velocities at the nodes are not parallel. For the field parallel (normal) to the heat current, the majority of unpaired states have $\bm v_F$ nearly normal (parallel) to the heat current [Figs.\:\ref{fig:nodes2D}(f),(g)] yielding a dominant twofold component [see the red curve in Fig.\:\ref{fig:nodes2D}(h)]. In contrast, the nodes on the flat part of the ellipse [in the case of Fig.\:\ref{fig:nodes2D}(b)] have a nearly identical direction of the Fermi velocity [Figs.\:\ref{fig:nodes2D}(d), and 3(e)], and contribute equally to the $\bm q$ for both $\bm H\| \bm q$ and $\bm H\perp\bm q$ [see the black curve in Fig.\:\ref{fig:nodes2D}(h)]. Thus our observation that a large fourfold component dominates over the twofold one showing clear minima near $\pm45^\circ$ eliminates the possibility of the nodal loops at the sharp edges of the FS (Fig.\:\ref{fig:nodes2D}(c)). It also argues against vertical line nodes [Fig.\:\ref{fig:nodes2D}(a)] as this order parameter structure would essentially average the two curves in Fig.\:\ref{fig:nodes2D}(h), leaving a large twofold anisotropy \cite{Vor10} not seen in our experiment.

\begin{figure}[t]
\begin{center}
\includegraphics[width=0.9\columnwidth]{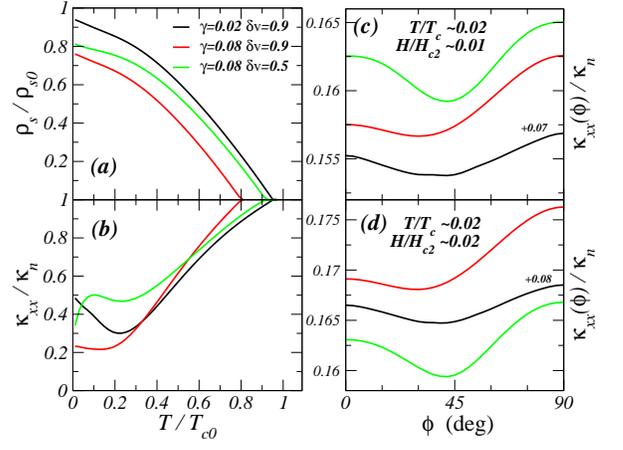}
\caption{
(Color online)
Results of the calculations for the 3D FSs with nodal loops. (a) The superfluid density remains linear down to low temperatures for different values of the scattering rate $\gamma$. (b) At the same time a large residual term in the thermal conductivity in zero field is observed. (c), (d) Anisotropy of thermal conductivity at low $T$ for several values of $H$. A large fourfold term with a minimum at 45$^\circ$ is observed  for cleaner material with sizable inter-band scattering (as well as for the dirtier case with small inter-band scattering).}
\label{fig:3dgap}
\end{center}
\end{figure}

To confirm that the nodal loops as sketched in Fig.\:\ref{fig:nodes2D}(b) consistently describe the bulk of the experimental measurements, we performed calculations for a three-dimensional (3D) FS given by $2mE_f = (k_xa)^2/(1-\mu\cos \frac{k_zc}{2}) + (k_ya)^2/(1+\mu\cos \frac{k_zc}{2})$.  For $\mu=0.5$ the FS closely mimics that obtained in the band structure calculations. We took the order parameter in the form $\Delta=\Delta_e Y(\phi, k_z)$, where $Y(\phi, k_z)= 1-|r(k_z)|+r(k_z)\cos 2\phi$, and $r(k_z)=-0.6\cos \frac{k_zc}{2}$. We also took the hole FS to be cylindrical with an isotropic gap $\Delta_h$, and self-consistently determined both $\Delta_e (T,\bm H)$ and $\Delta_h(T, \bm H)$ assuming a dominant interband pairing interaction. Figure\:\ref{fig:3dgap} shows that within this framework we can explain all the salient experimental features: linear-in-$T$ penetration depth, a large residual term in the thermal conductivity~\cite{Mish09}, and the dominant fourfold profile of $\kappa(\phi)$ under rotated field with a minimum along the 45$^\circ$ direction. We therefore believe that this is the most likely location of the gap nodes in BaFe$_2$(As$_{0.67}$P$_{0.33}$)$_2$.

Finally we discuss why such a nodal loop structure is realized. Experimentally, recent neutron scattering measurements suggest that the line nodes should create only a limited area of sign-reversal on a single FS \cite{Ish10}, which is consistent with the present nodal loop structure. The Raman scattering in Ba(Fe,Co)$_2$As$_2$ suggested a similar loop structure of gap minima \cite{Maz10}. A number of model calculations predict that the superconducting gap in the electron FSs can be more anisotropic than that in hole bands, and that nodes can appear in some parameter range \cite{Gra10,Kur09,Pla10,Chu09}.  In that case placing the nodes in the flat FS regions with low DOS does not significantly reduce the SC condensation energy.  In addition, it has been suggested that the nesting between the FS regions with the same orbital character is important for the pairing interaction.  We note that the change of the orbital character from $xy$ to $xz+yz$ occurs near the rim of the flat part of the electron pocket \cite{Ike11}, implying that the orbital character may favor the closed-loop nodes.  Thus the pairing interaction and low DOS stabilize the nodal loop structure with no serious reduction of $T_c$. A more detailed fully microscopic calculation is required to clarify the origin of the closed nodal loop structure, which is unique among unconventional superconductors.

In summary, we determined the nodal gap structure of BaFe$_2$(As$_{0.67}$P$_{0.33}$)$_2$ by angle-resolved thermal conductivity measurements.  We found distinct fourfold oscillation in $\kappa(\phi)$ with minima at [$\pm1$,$\pm1$,0] directions. The observed results are most consistent with the closed nodal loops at the flat part of the electron FS.

We thank A.\,V. Chubukov, P.\,J. Hirschfeld, H. Kontani, K. Kuroki, Y. Nagai, R. Prozorov and D.\,J. Scalapino for discussions. I.\,V. is supported in part by DOE Grant No. DE-FG02-08ER46492, and A.\,B.\,V. acknowledges support from NSF Grant No. DMR-0954342. I.\,V., A.\,B.\,V, and Y.\,M. thank the hospitality of KITP, made possible with support from NSF Grant No. PHY05-51164.

\end{document}